\documentclass[prl,showpacs,twocolumn, superscriptaddress]{revtex4}  
\usepackage{graphics}
\usepackage{amsmath}
\usepackage{hyperref}

\begin{document}

\title{Approaching the Heisenberg limit in an atom laser}

\author{M.~Jeppesen}
\affiliation{Australian Research Council Centre Of Excellence for Quantum-Atom Optics, Department of Physics, The Australian National University, Canberra, ACT 0200, Australia}
\email{matthew.jeppesen@anu.edu.au}
\homepage{http://www.acqao.org}

\author{J.~Dugu\'e}
\affiliation{Australian Research Council Centre Of Excellence for Quantum-Atom Optics, Department of Physics, The Australian National University, Canberra, ACT 0200, Australia}
\affiliation{Laboratoire Kastler-Brossel, 24 rue Lhomond, 75231 Paris Cedex 05, France}

\author{G.~R.~Dennis}
\author{M.~T.~Johnsson}
\author{C.~Figl}
\author{N.~P.~Robins}
\author{J.~D.~Close}
\affiliation{Australian Research Council Centre Of Excellence for Quantum-Atom Optics, Department of Physics, The Australian National University, Canberra, ACT 0200, Australia}

\begin{abstract}
We present experimental and theoretical results showing the improved beam quality and reduced divergence of an atom laser produced by an optical Raman transition, compared to one produced by an RF transition. We show that Raman outcoupling can eliminate the diverging lens effect that the condensate has on the outcoupled atoms. This substantially improves the beam quality of the atom laser, and the improvement may be greater than a factor of ten for experiments with tight trapping potentials. We show that Raman outcoupling can produce atom lasers whose quality is only limited by the wavefunction shape of the condensate that produces them, typically a factor of 1.3 above the Heisenberg limit.

\end{abstract}


\maketitle
Experiments in ultracold dilute atomic gases have had an enormous impact on physics. The realization of Bose-Einstein condensates (BECs), degenerate Fermi gases, BEC-BCS crossover systems, and many others have resulted in many fundamental insights and a wealth of new results in both experiment and theory.  One exciting system to emerge from this research is the atom laser, a highly coherent, directional beam of degenerate atoms, controllably released from a BEC~\cite{wiseman,mewes,bloch,gerbier,cennini,hagley,guerin, ottl:2005}.  
The atom lasers demonstrated so far have produced beams many orders of magnitude brighter than is possible with thermal atomic beams~\cite{robins:2006}.

\begin{figure}[b!]
\centerline{\scalebox{.7}{\includegraphics{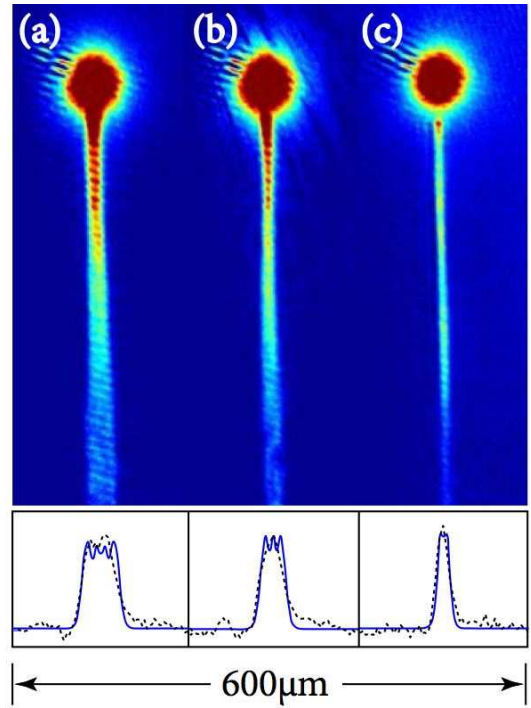}}}
\caption{(color online). Top: Sequence of atom laser beams showing the improved beam profile of a Raman atom laser. The atom laser beams were produced using RF (a) and Raman (b and c) transitions. The angle between the Raman beams (see Fig.~\ref{lasers}~(a)) was $\theta=30^{\circ}$ in (b) and $\theta=140^{\circ}$ in (c), corresponding to a kick of $0.5 \hbar k$ (0.3~cm/s) and $1.9 \hbar k$ (1.1~cm/s) respectively.  The outcoupling rate differs between each atom laser.
Below: Comparison of experimental (dashed) and theoretical (solid) beam profiles $500~\mu$m below the BEC.  The height of each theoretical curve has been scaled to match experimental data.}
\label{threelasers}
\end{figure}

Atom laser beams show great promise for studies of fundamental physics and in high precision measurements~\cite{kasevich}. In the future, it will be possible to produce quadrature squeezing in atoms lasers, to use atom lasers to produce correlations and entanglement between massive particles~\cite{haine:2006}, as well as high precision interferometers both on earth and in space~\cite{le-coq:2006}. For all these it will be crucial to develop atom lasers with output modes that as clean as possible in amplitude and phase, to allow stable modematching, just as it was crucial for optical lasers. The beam quality factor $M^2$, introduced for atom lasers by J.-F. Riou \textit{et al.}~\cite{riou,siegman}, is a measure of how far the beam deviates from the Heisenberg limit, and is defined by
\begin{equation}
    M^2 = \frac{2}{\hbar} \Delta x \Delta p_x, \label{M2-def}
\end{equation}
where $\Delta x$ is the beam width, measured at the waist, and $\Delta p_x$ is the transverse momentum spread. An ideal (Gaussian) beam would therefore have $M^2 =1$ along both its principal transverse axes. A number of experimental works have shown that the beam quality of an atom laser is strongly affected by the interaction of the outcoupled atoms with the BEC from which it is produced~\cite{kohl,busch,riou,le-coq, ottl:2006}. As the atoms fall through the condensate, the repulsive interaction acts as a diverging lens to the outcoupled atoms.
This leads to a divergence in the atom laser beam and (because the BEC is a non-ideal lens) a poor quality transverse beam profile.  Such behavior may cause problems in mode matching the atom laser beam to another atom laser, a cavity or to a waveguide. Experiments on atom lasers in waveguides have produced beams with improved spatial profile~\cite{guerin}. However, precision measurements with atom interferometry are likely to require propagation in free space, to avoid introducing noise from the fluctuations in the waveguide itself~\cite{le-coq:2006}.

In a recent Letter~\cite{riou}, it was shown that the quality of a free space atom laser is improved by outcoupling from the base of the condensate. Our scheme, however, enables the production of a high quality atom laser while outcoupling from the center of the condensate. This is desirable for a number of reasons: First, because the classical noise level is determined by the outcoupling Rabi frequency, then outcoupling from the center, where the density is greatest, gives the highest possible output flux for a given classical noise level~\cite{robins:2005}.  Second, outcoupling from the center allows the longest operating time (for a quasicontinuous atom laser) since the condensate can be drained completely.  Third, outcoupling from the center minimizes the sensitivity of the output coupling to condensate excitations or external fluctuations.

In a recent Letter~\cite{robins:2006}, we have demonstrated a continuously outcoupled atom laser where the output coupler is a coherent multi-photon (Raman) transition~\cite{hagley,ruostekoski}. In this scheme, the atoms receive a momentum kick from the absorption and emission of photons. They leave the condensate more quickly, so that adverse effects due to the mean-field repulsion from the condensate are reduced. In this Letter, we report measurements of a substantial improvement in the beam quality $M^2$ using this outcoupling. In Fig.~\ref{threelasers}, we show absorption images of atom laser beams outcoupled from the center of a BEC with (a) negligible momentum kick, (b) a kick of 0.3~cm/s, and 1.1~cm/s (c). As the kick increases, the divergence is reduced and the beam profile improved.


In our experiment, we create $^{87}$Rb BECs of \mbox{$5 \times 10^5$~atoms} in the $|F=1, m_F=-1 \rangle $ state via standard runaway evaporation of laser cooled atoms.  We use a highly stable, water cooled QUIC magnetic trap (axial frequency $\omega_{y} = 2 \pi \times 12$~Hz and radial frequency $\omega_{\rho} = 2 \pi \times 128$~Hz, with a bias field of $B_0=2$~G). We control drifts in the magnetic bias by using high stability power supplies and water cooling.  This stability allows us to precisely and repeatably address the condensate.  

\begin{figure}[t]
\centerline{\scalebox{.33}{\includegraphics{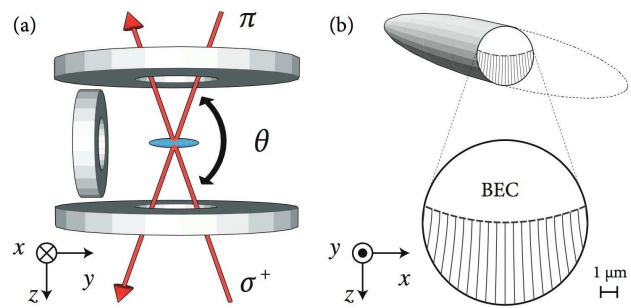}}}
\caption{(color online) (a) Experimental schematic (not to scale) showing the BEC, Raman lasers, and trapping coils. (b) Cross section along the two strong axes of the magnetic trap, showing the BEC, outcoupling surface, and atom laser trajectories. Note that the field of view in (b) is rotated $90^{\circ}$ with respect to (a).} \label{lasers}
\end{figure}

We produce the atom laser by transferring the atoms to the untrapped $|F=1, m_F=0 \rangle$ state and letting them fall under gravity. To outcouple atoms with negligible momentum kick we induce spin flips via an RF field of a frequency corresponding to the Zeeman shift in the center of the condensate. Alternatively, we induce the spin flips via an optical Raman transition. The setup is shown in Fig.~\ref{lasers}~(a). Two optical Raman beams, separated by an angle $\theta$, propagate in the plane of gravity and the magnetic trap bias field. The momentum transfer to the atoms through absorption and emission of the photons is ${ 2\hbar k \sin(\theta/2)}$, with $k$ the wave number of the laser beams.  The Raman laser beams are produced from one 700~mW diode laser. We can turn the laser power on or off in less than 200~ns using a fast switching AOM in a double pass configuration. After the switching AOM, the light is split and sent through two separate AOMs, again each in a double pass configuration. The frequency difference between the AOMs corresponds to the Zeeman plus kinetic energy difference between the initial and final states of the two-photon Raman transition.  We stabilize the frequency difference by  running the 80~MHz function generators driving the AOMs from a single oscillator.  The beams are then coupled via single mode, polarization maintaining optical fibers directly to the BEC through a collimating lens and waveplate, providing a maximum intensity of 2500~mW/cm$^2$ per beam at the BEC. The polarization of the beams is optimized to achieve maximum outcoupling with a downward kick and corresponds to $\pi$ polarization for the upper beam and $\sigma^+$ for the lower beam.

\begin{figure}[t]
\centerline{\scalebox{.33}{\includegraphics{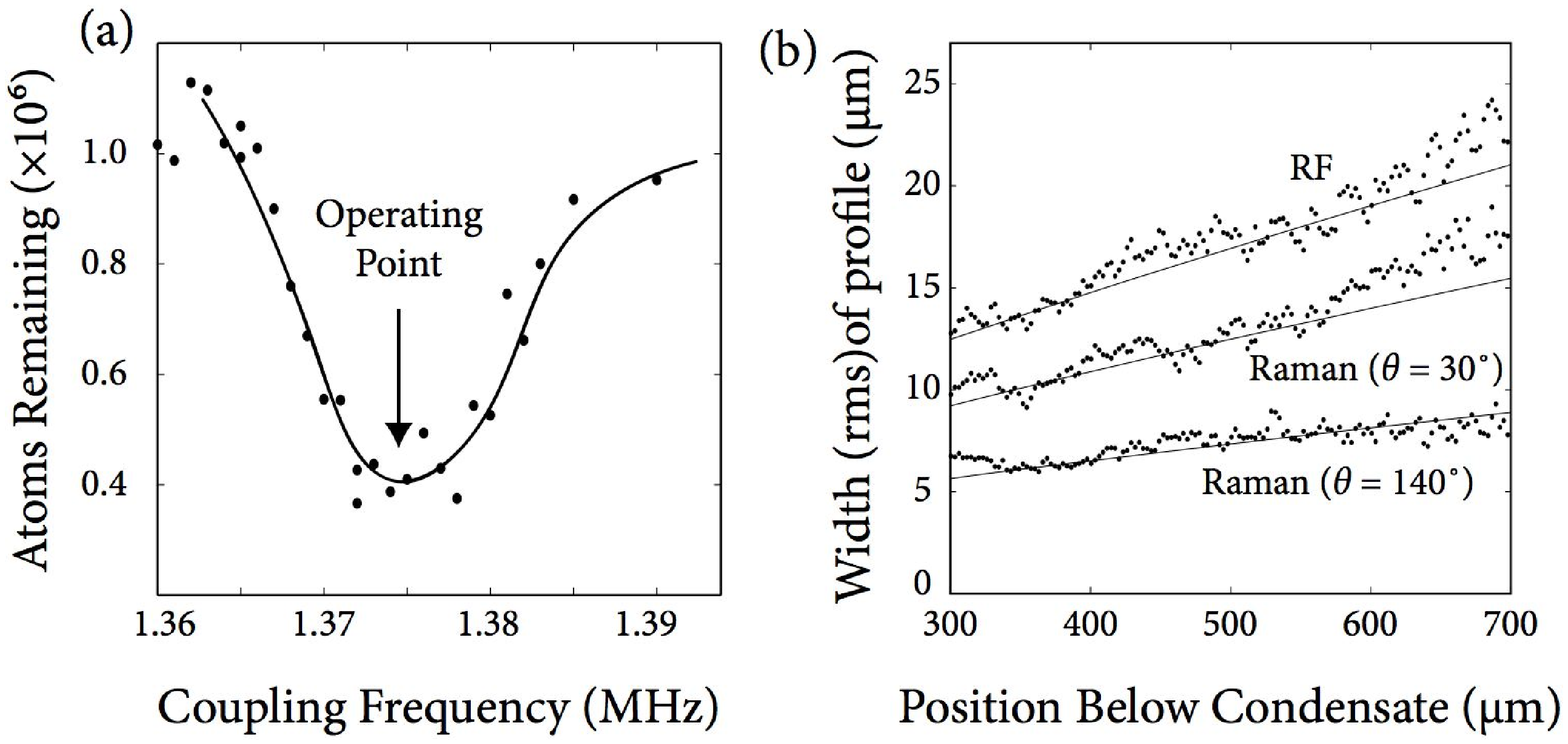}}}
\caption{(a) Output coupling spectroscopy showing the operating point at the center of the BEC, solid curve to guide the eye.  
(b) The rms beam width for RF and Raman atom lasers. The dots represent experimental measurements and the solid curves our theoretical predictions.
}\label{spectroscopy}

\end{figure}

The outcoupling resonance is set to the center of the BEC for both RF and Raman outcoupling, as shown in Fig.~\ref{lasers}~(b).  This point is found by performing spectroscopy on the BEC using 100~ms of weak output coupling at varying RF frequencies, and measuring the number of atoms remaining in the condensate after the output coupling time \cite{bloch}.  A typical calibration curve is shown in Fig.~\ref{spectroscopy}~(a), in this case for RF outcoupling. We operate both RF and Raman output couplers at the point of maximum outcoupling rate. We further check this frequency by ensuring that a continuous beam can still be produced when the initial condensate is very small, which can only happen when outcoupling from the center. 

We observe the system using standard absorption imaging along the $y$ (weak trapping) direction, on the $F = 2 \rightarrow F'=3$ transition, with a 200~$\mu$s pulse of repumping light ($F = 1 \rightarrow F'=2$) 1~ms prior to imaging. From these images we are able to extract the rms width of the atom laser as a function of fall distance (see Fig.~\ref{spectroscopy}~(b)), which we use to calculate $M^2$ (details below). 

To model the system, we use a two-step method following \cite{riou}. Inside the condensate, we use the WKB approximation, by integrating the phase along the classical trajectories of atoms moving in the Thomas-Fermi potential of the condensate (an inverted paraboloid)~\cite{busch}.  After this, we propagate the atom laser wavefunction using a Kirchoff-Fresnel diffraction integral over the surface of the condensate: 
\begin{equation}
   \psi(\mathbf{r}) = \int_S
d \mathbf{S'} \cdot [ G \, \nabla' \, \psi - \psi \, \nabla' \, G], \label{green's}
\end{equation}
where $G = G(\mathbf{r}, \mathbf{r'})$ is the Green's function for the Hamiltonian in the gravitational potential $V(\mathbf{r}) = - m g z$ \cite{borde}. Therefore, the model includes only interactions between condensate atoms and beam atoms; interactions between atoms within the beam are ignored. The integral in Eq.~(\ref{green's}) is formally a two dimensional surface integral over the whole condensate. However for simplicity, following \cite{riou}, we neglect divergence in the weak trapping axis and only consider cross sections in the plane of the strong trapping axes, and so the integral becomes one dimensional. A 3D wavefunction is built up by calculating the atom laser in a series of planes along the weak trapping axis. 
 
We ignore the effects of the magnetic field on the atom laser. The atom laser state $| F = 1, m_F = 0 \rangle$ is unaffected to first order by the magnetic field, but is weakly anti-trapped due to the second order Zeeman effect, with an effective  trapping frequency of $\omega_{\text{2nd}} = 2 \pi \times 2.6$~Hz. The transverse position of an atom in such a potential is 
\begin{equation}
    x(t) = x_0 \cosh(\omega_{\text{2nd}} t) \approx x_0( 1 + \omega_{\text{2nd}}^2 t^2/2).
\end{equation}
For the 1~mm (14~ms) propagation we consider here the transverse position is affected by less than 3\%. We also ignore the AC Stark effect of the Raman beams on the atom laser, because the intensity of the beams does not change significantly over the 1~mm propagation.

We have checked the validity of this model against a solution of the full 3D Gross-Pitaesvskii (GP) equation, including beam-beam interactions. To find the atom laser wavefunction at large distances below the condensate (up to 1~mm), we transfer the GP model to a freely falling frame once the atom laser wavefunction has reached steady state. The details of the calculation will be the basis of a future publication. The two models give good agreement.


Calculating the quality factor $M^2$ of the atom laser directly from Eq.~(\ref{M2-def}) requires measurement of the beam width at the waist $\Delta x_0$. Because the BEC acts as a diverging lens on the atom laser, the beam waist is \emph{virtual} and located above the BEC, and so it is not possible to measure the beam quality $M^2$ using Eq.~\ref{M2-def} only. For our simulations, $M^2$ is calculated equivalently from the wavefunction $\psi(x,y,z)$ at some height $z$ below the BEC in which the atom laser has reached the paraxial regime:
\begin{equation}
    (M^2/2)^2 = ( \Delta x(z))^2 (\Delta k_x(z))^2 - C(z)^2,   \label{M2-alt-def}
\end{equation}
where $\Delta x(z)$ is beam width and  $C(z)$ is the curvature-beam width product~\cite{riou:phd}:
   \begin{equation}
C(z) = \frac{i}{2} \int_{-\infty}^{\infty} x \left( \psi \frac{\partial \psi^*}{\partial x} - \psi^* \frac{\partial \psi}{\partial x} \right) dx.
    \end{equation}
%
In practice it is difficult to measure the wavefunction phase, and hence $C(z)$. However the beam width, in the paraxial regime, obeys:
  \begin{equation}
\Delta (x (t))^2 = (\Delta x_0)^2 + (\Delta v_x)^2 (t-t_w)^2, \label{width_as_fn_of_t}
    \end{equation}
where $t_w$ is the time when the beam is at its waist, and $\Delta x_0$ is the beam waist. In principle $M^2$ may be determined simply from measurements of the beam width at different heights. In our experiment, we can only measure the beam width in the far field, at distances greater than 300~$\mu$m below the condensate (observation at distances less than 300~$\mu$m are prevented by the condensate expansion after trap switchoff.) In the far field the second term of Eq.~\ref{width_as_fn_of_t}\ dominates, and so only the velocity spread can be measured. Therefore we calculate $\Delta x_0$ and $t_w$ from the model, $t_w = m \, C(z)/ (\hbar \Delta k_x^2)$, with $t_w$ negative since the waist is virtual and located above the BEC. We then fit to the experimental data to find $\Delta v_x$. 

\begin{figure}[t]
\centerline{\scalebox{.45}{\includegraphics{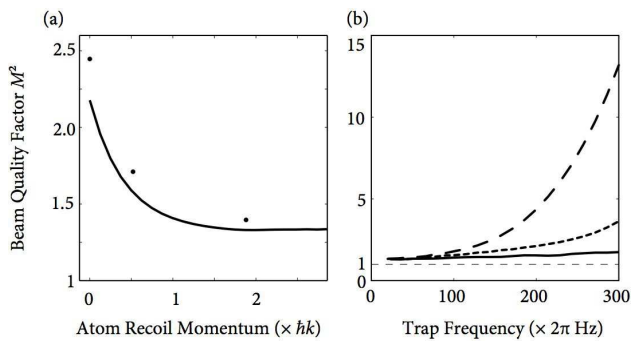}}}
\caption{(a) Calculated quality factor $M^2$ of an atom laser. The dots are the experimental measurements, and the solid line our theoretical predictions. (b) $M^2$ as a function of trapping frequency for an RF atom laser (dashed line), a kick of $0.5 \hbar k$ (0.3~cm/s) (dotted line), and $2 \hbar k$ (1.1~cm/s) (solid line). The condensate number was $N = 5 \times 10^5$ atoms, and the aspect ratio $\omega_{\rho} / \omega_y$ was 10.}   \label{M2} 
\end{figure}

In Fig.~\ref{M2}, we present the theoretical and experimental results. We find that as the kick increases, the beam quality is improved and the divergence is reduced. For our parameters, we find that for an RF atom laser $M^2 = 2.2$, and for a Raman atom laser $M^2 = 1.4$ with the maximum two photon kick. As the kick increases, $M^2$ continues to improve, and approaches but does not reach the Heisenberg limit of one. It asymptotes to a limit slightly above that, which for our parameters is equal to 1.3. In this regime of large kick, the interaction of the outcoupled atoms with the condensate becomes negligible, and the transverse atom laser wavefunction is approximately the free space evolution of the condensate wavefunction (along the outcoupling surface). It is therefore limited by the non-ideal (non-Gaussian) condensate wavefunction itself. We calculate the product $\Delta x \Delta p_x$ for the condensate wavefunction (taken through the central horizontal plane of the condensate) to be 1.3. We have therefore improved the beam quality $M^2$ by 50 percent, down to a factor of 1.4 above the Heisenberg limit. In addition, our simulations show that (using the same maximum two photon kick) it is possible to reach the condensate limit even for much tighter trapping potentials. In Fig~\ref{M2} (b), we show the results of simulations for increasing trap frequencies, up to $\omega = 2 \pi \times 300$ Hz. As the trap frequency increases, the $M^2$ worsens, up to $M^2 = 14$  for RF outcoupling from a $2 \pi \times 300$~Hz trap. For the maximum Raman two photon kick, the increase is only to $M^2 = 1.7$ for the same $2 \pi \times 300$~Hz trap. Only for traps of less than $2 \pi \times 50$~Hz is the beam quality of an RF atom laser within 5 percent of that of a Raman atom laser. 

With higher order Raman transitions \cite{kozuma}, it will be possible to reach the condensate limit even for experiments with traps of several kilohertz. It will also be possible to reach the Heisenberg limit by completely removing the atomic interaction, for example by using a Feschbach resonance. Using Raman lasers phase locked to the 6.8~GHz hyperfine splitting will prevent populating the anti-trapped state, and produce a truly two state atom laser~\cite{ottl:2006,dugue:2007}. Such lasers, combined with the high quality transverse mode of Raman atom lasers, could be used in a continuous version of the atomic Mach-Zehnder Bragg interferometer~\cite{torii:2000}, and in the development of atomic local oscillators.

We thank Ruth Mills for useful discussions. CF acknowledges funding from the Alexander von Humboldt foundation. This work was financially supported by the Australian Research Council Centre of Excellence program. Numerical simulations were done at the APAC National Supercomputing Facility.

\bibliography{bibliography}

\begin{thebibliography}{25}
\expandafter\ifx\csname natexlab\endcsname\relax\def\natexlab#1{#1}\fi
\expandafter\ifx\csname bibnamefont\endcsname\relax
  \def\bibnamefont#1{#1}\fi
\expandafter\ifx\csname bibfnamefont\endcsname\relax
  \def\bibfnamefont#1{#1}\fi
\expandafter\ifx\csname citenamefont\endcsname\relax
  \def\citenamefont#1{#1}\fi
\expandafter\ifx\csname url\endcsname\relax
  \def\url#1{\texttt{#1}}\fi
\expandafter\ifx\csname urlprefix\endcsname\relax\def\urlprefix{URL }\fi
\providecommand{\bibinfo}[2]{#2}
\providecommand{\eprint}[2][]{\url{#2}}

\bibitem[{\citenamefont{Wiseman}(1997)}]{wiseman}
\bibinfo{author}{\bibfnamefont{H.~M.} \bibnamefont{Wiseman}},
  \bibinfo{journal}{Phys. Rev. A} \textbf{\bibinfo{volume}{56}},
  \bibinfo{pages}{2068} (\bibinfo{year}{1997}).

\bibitem[{\citenamefont{Mewes et~al.}(1997)\citenamefont{Mewes, Andrews, Kurn,
  Durfee, Townsend, and Ketterle}}]{mewes}
\bibinfo{author}{\bibfnamefont{M.-O.} \bibnamefont{Mewes}},
  \bibinfo{author}{\bibfnamefont{M.~R.} \bibnamefont{Andrews}},
  \bibinfo{author}{\bibfnamefont{D.~M.} \bibnamefont{Kurn}},
  \bibinfo{author}{\bibfnamefont{D.~S.} \bibnamefont{Durfee}},
  \bibinfo{author}{\bibfnamefont{C.~G.} \bibnamefont{Townsend}},
  \bibnamefont{and} \bibinfo{author}{\bibfnamefont{W.}~\bibnamefont{Ketterle}},
  \bibinfo{journal}{Phys. Rev. Lett.} \textbf{\bibinfo{volume}{78}},
  \bibinfo{pages}{582} (\bibinfo{year}{1997}).

\bibitem[{\citenamefont{Bloch et~al.}(1999)\citenamefont{Bloch, H{\"a}nsch, and
  Esslinger}}]{bloch}
\bibinfo{author}{\bibfnamefont{I.}~\bibnamefont{Bloch}},
  \bibinfo{author}{\bibfnamefont{T.~W.} \bibnamefont{H{\"a}nsch}},
  \bibnamefont{and}
  \bibinfo{author}{\bibfnamefont{T.}~\bibnamefont{Esslinger}},
  \bibinfo{journal}{Phys. Rev. Lett.} \textbf{\bibinfo{volume}{82}},
  \bibinfo{pages}{3008} (\bibinfo{year}{1999}).

\bibitem[{\citenamefont{Gerbier et~al.}(2001)\citenamefont{Gerbier, Bouyer, and
  Aspect}}]{gerbier}
\bibinfo{author}{\bibfnamefont{F.}~\bibnamefont{Gerbier}},
  \bibinfo{author}{\bibfnamefont{P.}~\bibnamefont{Bouyer}}, \bibnamefont{and}
  \bibinfo{author}{\bibfnamefont{A.}~\bibnamefont{Aspect}},
  \bibinfo{journal}{Phys. Rev. Lett.} \textbf{\bibinfo{volume}{86}}
  (\bibinfo{year}{2001}).

\bibitem[{\citenamefont{{Cennini} et~al.}(2003)\citenamefont{{Cennini}, {Ritt},
  {Geckeler}, and {Weitz}}}]{cennini}
\bibinfo{author}{\bibfnamefont{G.}~\bibnamefont{{Cennini}}},
  \bibinfo{author}{\bibfnamefont{G.}~\bibnamefont{{Ritt}}},
  \bibinfo{author}{\bibfnamefont{C.}~\bibnamefont{{Geckeler}}},
  \bibnamefont{and} \bibinfo{author}{\bibfnamefont{M.}~\bibnamefont{{Weitz}}},
  \bibinfo{journal}{Phys. Rev. Lett.} \textbf{\bibinfo{volume}{91}},
  \bibinfo{pages}{240408} (\bibinfo{year}{2003}).

\bibitem[{\citenamefont{Hagley et~al.}(1999)\citenamefont{Hagley, Deng, Kozuma,
  Wren, Helmerson, Rolston, and Phillips}}]{hagley}
\bibinfo{author}{\bibfnamefont{E.~W.} \bibnamefont{Hagley}},
  \bibinfo{author}{\bibfnamefont{L.}~\bibnamefont{Deng}},
  \bibinfo{author}{\bibfnamefont{M.}~\bibnamefont{Kozuma}},
  \bibinfo{author}{\bibfnamefont{J.}~\bibnamefont{Wren}},
  \bibinfo{author}{\bibfnamefont{K.}~\bibnamefont{Helmerson}},
  \bibinfo{author}{\bibfnamefont{S.~L.} \bibnamefont{Rolston}},
  \bibnamefont{and} \bibinfo{author}{\bibfnamefont{W.~D.}
  \bibnamefont{Phillips}}, \bibinfo{journal}{Science}
  \textbf{\bibinfo{volume}{283}}, \bibinfo{pages}{1706} (\bibinfo{year}{1999}).

\bibitem[{\citenamefont{{Guerin} et~al.}(2006)\citenamefont{{Guerin}, {Riou},
  {Gaebler}, {Josse}, {Bouyer}, and {Aspect}}}]{guerin}
\bibinfo{author}{\bibfnamefont{W.}~\bibnamefont{{Guerin}}},
  \bibinfo{author}{\bibfnamefont{J.-F.} \bibnamefont{{Riou}}},
  \bibinfo{author}{\bibfnamefont{J.~P.} \bibnamefont{{Gaebler}}},
  \bibinfo{author}{\bibfnamefont{V.}~\bibnamefont{{Josse}}},
  \bibinfo{author}{\bibfnamefont{P.}~\bibnamefont{{Bouyer}}}, \bibnamefont{and}
  \bibinfo{author}{\bibfnamefont{A.}~\bibnamefont{{Aspect}}},
  \bibinfo{journal}{Phys. Rev. Lett.} \textbf{\bibinfo{volume}{97}},
  \bibinfo{pages}{200402} (\bibinfo{year}{2006}).

\bibitem[{\citenamefont{\"Ottl et~al.}(2005)\citenamefont{\"Ottl, Ritter, and
  K\"ohl}}]{ottl:2005}
\bibinfo{author}{\bibfnamefont{A.}~\bibnamefont{\"Ottl}},
  \bibinfo{author}{\bibfnamefont{S.}~\bibnamefont{Ritter}}, \bibnamefont{and}
  \bibinfo{author}{\bibfnamefont{T.}~\bibnamefont{K\"ohl},
  \bibfnamefont{M.~Esslinger}}, \bibinfo{journal}{Phys. Rev. Lett.}
  \textbf{\bibinfo{volume}{95}}, \bibinfo{pages}{090404}
  (\bibinfo{year}{2005}).

\bibitem[{\citenamefont{{Robins} et~al.}(2006)\citenamefont{{Robins}, {Figl},
  {Haine}, {Morrison}, {Jeppesen}, {Hope}, and {Close}}}]{robins:2006}
\bibinfo{author}{\bibfnamefont{N.~P.} \bibnamefont{{Robins}}},
  \bibinfo{author}{\bibfnamefont{C.}~\bibnamefont{{Figl}}},
  \bibinfo{author}{\bibfnamefont{S.~A.} \bibnamefont{{Haine}}},
  \bibinfo{author}{\bibfnamefont{A.~K.} \bibnamefont{{Morrison}}},
  \bibinfo{author}{\bibfnamefont{M.}~\bibnamefont{{Jeppesen}}},
  \bibinfo{author}{\bibfnamefont{J.~J.} \bibnamefont{{Hope}}},
  \bibnamefont{and} \bibinfo{author}{\bibfnamefont{J.~D.}
  \bibnamefont{{Close}}}, \bibinfo{journal}{Phys. Rev. Lett.}
  \textbf{\bibinfo{volume}{96}}, \bibinfo{pages}{140403}
  (\bibinfo{year}{2006}).

\bibitem[{\citenamefont{Kasevich}(2002)}]{kasevich}
\bibinfo{author}{\bibfnamefont{M.~A.} \bibnamefont{Kasevich}},
  \bibinfo{journal}{Science} \textbf{\bibinfo{volume}{298}},
  \bibinfo{pages}{1363} (\bibinfo{year}{2002}).

\bibitem[{\citenamefont{{Haine} et~al.}(2006)\citenamefont{{Haine}, {Olsen},
  and {Hope}}}]{haine:2006}
\bibinfo{author}{\bibfnamefont{S.~A.} \bibnamefont{{Haine}}},
  \bibinfo{author}{\bibfnamefont{M.~K.} \bibnamefont{{Olsen}}},
  \bibnamefont{and} \bibinfo{author}{\bibfnamefont{J.~J.}
  \bibnamefont{{Hope}}}, \bibinfo{journal}{Phys. Rev. Lett.}
  \textbf{\bibinfo{volume}{96}}, \bibinfo{pages}{133601}
  (\bibinfo{year}{2006}).

\bibitem[{\citenamefont{{Le Coq} et~al.}(2006)\citenamefont{{Le Coq}, {Retter},
  {Richard}, {Aspect}, and {Bouyer}}}]{le-coq:2006}
\bibinfo{author}{\bibfnamefont{Y.}~\bibnamefont{{Le Coq}}},
  \bibinfo{author}{\bibfnamefont{J.~A.} \bibnamefont{{Retter}}},
  \bibinfo{author}{\bibfnamefont{S.}~\bibnamefont{{Richard}}},
  \bibinfo{author}{\bibfnamefont{A.}~\bibnamefont{{Aspect}}}, \bibnamefont{and}
  \bibinfo{author}{\bibfnamefont{P.}~\bibnamefont{{Bouyer}}},
  \bibinfo{journal}{App. Phys. B} \textbf{\bibinfo{volume}{84}},
  \bibinfo{pages}{627} (\bibinfo{year}{2006}).

\bibitem[{\citenamefont{Riou et~al.}(2006)\citenamefont{Riou, Guerin, Coq,
  Fauquembergue, Josse, Bouyer, and Aspect}}]{riou}
\bibinfo{author}{\bibfnamefont{J.-F.} \bibnamefont{Riou}},
  \bibinfo{author}{\bibfnamefont{W.}~\bibnamefont{Guerin}},
  \bibinfo{author}{\bibfnamefont{Y.~L.} \bibnamefont{Coq}},
  \bibinfo{author}{\bibfnamefont{M.}~\bibnamefont{Fauquembergue}},
  \bibinfo{author}{\bibfnamefont{V.}~\bibnamefont{Josse}},
  \bibinfo{author}{\bibfnamefont{P.}~\bibnamefont{Bouyer}}, \bibnamefont{and}
  \bibinfo{author}{\bibfnamefont{A.}~\bibnamefont{Aspect}},
  \bibinfo{journal}{Phys. Rev. Lett.} \textbf{\bibinfo{volume}{96}},
  \bibinfo{pages}{070404} (\bibinfo{year}{2006}).

\bibitem[{\citenamefont{{Siegman}}(1991)}]{siegman}
\bibinfo{author}{\bibfnamefont{A.~E.} \bibnamefont{{Siegman}}},
  \bibinfo{journal}{IEEE. J. Quantum Electron.} \textbf{\bibinfo{volume}{27}},
  \bibinfo{pages}{1146} (\bibinfo{year}{1991}).

\bibitem[{\citenamefont{{K{\"o}hl} et~al.}(2005)\citenamefont{{K{\"o}hl},
  {Busch}, {M{\o}lmer}, {H{\"a}nsch}, and {Esslinger}}}]{kohl}
\bibinfo{author}{\bibfnamefont{M.}~\bibnamefont{{K{\"o}hl}}},
  \bibinfo{author}{\bibfnamefont{T.}~\bibnamefont{{Busch}}},
  \bibinfo{author}{\bibfnamefont{K.}~\bibnamefont{{M{\o}lmer}}},
  \bibinfo{author}{\bibfnamefont{T.~W.} \bibnamefont{{H{\"a}nsch}}},
  \bibnamefont{and}
  \bibinfo{author}{\bibfnamefont{T.}~\bibnamefont{{Esslinger}}},
  \bibinfo{journal}{\pra} \textbf{\bibinfo{volume}{72}},
  \bibinfo{pages}{063618} (\bibinfo{year}{2005}).

\bibitem[{\citenamefont{Busch et~al.}(2002)\citenamefont{Busch, K{\"o}hl,
  Esslinger, and M{\o}lmer}}]{busch}
\bibinfo{author}{\bibfnamefont{T.}~\bibnamefont{Busch}},
  \bibinfo{author}{\bibfnamefont{M.}~\bibnamefont{K{\"o}hl}},
  \bibinfo{author}{\bibfnamefont{T.}~\bibnamefont{Esslinger}},
  \bibnamefont{and}
  \bibinfo{author}{\bibfnamefont{K.}~\bibnamefont{M{\o}lmer}},
  \bibinfo{journal}{Phys. Rev. A} \textbf{\bibinfo{volume}{65}},
  \bibinfo{pages}{043615} (\bibinfo{year}{2002}).

\bibitem[{\citenamefont{Coq et~al.}(2001)\citenamefont{Coq, Thywissen,
  Rangwala, Gerbier, Richard, Delannoy, Bouyer, and Aspect}}]{le-coq}
\bibinfo{author}{\bibfnamefont{Y.~L.} \bibnamefont{Coq}},
  \bibinfo{author}{\bibfnamefont{J.~H.} \bibnamefont{Thywissen}},
  \bibinfo{author}{\bibfnamefont{S.~A.} \bibnamefont{Rangwala}},
  \bibinfo{author}{\bibfnamefont{F.}~\bibnamefont{Gerbier}},
  \bibinfo{author}{\bibfnamefont{S.}~\bibnamefont{Richard}},
  \bibinfo{author}{\bibfnamefont{G.}~\bibnamefont{Delannoy}},
  \bibinfo{author}{\bibfnamefont{P.}~\bibnamefont{Bouyer}}, \bibnamefont{and}
  \bibinfo{author}{\bibfnamefont{A.}~\bibnamefont{Aspect}},
  \bibinfo{journal}{Phys. Rev. Lett.} \textbf{\bibinfo{volume}{87}},
  \bibinfo{pages}{170403} (\bibinfo{year}{2001}).

\bibitem[{\citenamefont{\"Ottl et~al.}(2006)\citenamefont{\"Ottl, Ritter, Kohl,
  and Esslinger}}]{ottl:2006}
\bibinfo{author}{\bibfnamefont{A.}~\bibnamefont{\"Ottl}},
  \bibinfo{author}{\bibfnamefont{S.}~\bibnamefont{Ritter}},
  \bibinfo{author}{\bibfnamefont{M.}~\bibnamefont{Kohl}}, \bibnamefont{and}
  \bibinfo{author}{\bibfnamefont{T.}~\bibnamefont{Esslinger}},
  \bibinfo{journal}{Rev. of Sci. Instrum.} \textbf{\bibinfo{volume}{77}},
  \bibinfo{pages}{063118} (\bibinfo{year}{2006}).

\bibitem[{\citenamefont{{Robins} et~al.}(2005)\citenamefont{{Robins},
  {Morrison}, {Hope}, and {Close}}}]{robins:2005}
\bibinfo{author}{\bibfnamefont{N.~P.} \bibnamefont{{Robins}}},
  \bibinfo{author}{\bibfnamefont{A.~K.} \bibnamefont{{Morrison}}},
  \bibinfo{author}{\bibfnamefont{J.~J.} \bibnamefont{{Hope}}},
  \bibnamefont{and} \bibinfo{author}{\bibfnamefont{J.~D.}
  \bibnamefont{{Close}}}, \bibinfo{journal}{\pra}
  \textbf{\bibinfo{volume}{72}}, \bibinfo{pages}{031606}
  (\bibinfo{year}{2005}).

\bibitem[{\citenamefont{{Ruostekoski} et~al.}(2003)\citenamefont{{Ruostekoski},
  {Gasenzer}, and {Hutchinson}}}]{ruostekoski}
\bibinfo{author}{\bibfnamefont{J.}~\bibnamefont{{Ruostekoski}}},
  \bibinfo{author}{\bibfnamefont{T.}~\bibnamefont{{Gasenzer}}},
  \bibnamefont{and}
  \bibinfo{author}{\bibfnamefont{D.}~\bibnamefont{{Hutchinson}}},
  \bibinfo{journal}{\pra} \textbf{\bibinfo{volume}{68}},
  \bibinfo{pages}{011604} (\bibinfo{year}{2003}).

\bibitem[{\citenamefont{Bord{\'e}}(2001)}]{borde}
\bibinfo{author}{\bibfnamefont{C.~J.} \bibnamefont{Bord{\'e}}},
  \bibinfo{journal}{C. R. Acad. Sci. Paris} \textbf{\bibinfo{volume}{4}},
  \bibinfo{pages}{509} (\bibinfo{year}{2001}).

\bibitem[{\citenamefont{{Riou}}(2006)}]{riou:phd}
\bibinfo{author}{\bibfnamefont{J.-F.} \bibnamefont{{Riou}}}, Ph.D. thesis,
  \bibinfo{school}{Institut D'Optique} (\bibinfo{year}{2006}).

\bibitem[{\citenamefont{{Kozuma} et~al.}(1999)\citenamefont{{Kozuma}, {Deng},
  {Hagley}, {Wen}, {Lutwak}, {Helmerson}, {Rolston}, and {Phillips}}}]{kozuma}
\bibinfo{author}{\bibfnamefont{M.}~\bibnamefont{{Kozuma}}},
  \bibinfo{author}{\bibfnamefont{L.}~\bibnamefont{{Deng}}},
  \bibinfo{author}{\bibfnamefont{E.~W.} \bibnamefont{{Hagley}}},
  \bibinfo{author}{\bibfnamefont{J.}~\bibnamefont{{Wen}}},
  \bibinfo{author}{\bibfnamefont{R.}~\bibnamefont{{Lutwak}}},
  \bibinfo{author}{\bibfnamefont{K.}~\bibnamefont{{Helmerson}}},
  \bibinfo{author}{\bibfnamefont{S.~L.} \bibnamefont{{Rolston}}},
  \bibnamefont{and} \bibinfo{author}{\bibfnamefont{W.~D.}
  \bibnamefont{{Phillips}}}, \bibinfo{journal}{Phys. Rev. Lett.}
  \textbf{\bibinfo{volume}{82}}, \bibinfo{pages}{871} (\bibinfo{year}{1999}).

\bibitem[{\citenamefont{{Dugu{\'e}} et~al.}(2007)\citenamefont{{Dugu{\'e}},
  {Robins}, {Figl}, {Jeppesen}, {Summers}, {Johnsson}, {Hope}, and
  {Close}}}]{dugue:2007}
\bibinfo{author}{\bibfnamefont{J.}~\bibnamefont{{Dugu{\'e}}}},
  \bibinfo{author}{\bibfnamefont{N.~P.} \bibnamefont{{Robins}}},
  \bibinfo{author}{\bibfnamefont{C.}~\bibnamefont{{Figl}}},
  \bibinfo{author}{\bibfnamefont{M.}~\bibnamefont{{Jeppesen}}},
  \bibinfo{author}{\bibfnamefont{P.}~\bibnamefont{{Summers}}},
  \bibinfo{author}{\bibfnamefont{M.~T.} \bibnamefont{{Johnsson}}},
  \bibinfo{author}{\bibfnamefont{J.~J.} \bibnamefont{{Hope}}},
  \bibnamefont{and} \bibinfo{author}{\bibfnamefont{J.~D.}
  \bibnamefont{{Close}}}, \bibinfo{journal}{\pra}
  \textbf{\bibinfo{volume}{75}}, \bibinfo{pages}{053602}
  (\bibinfo{year}{2007}).

\bibitem[{\citenamefont{{Torii} et~al.}(2000)\citenamefont{{Torii}, {Suzuki},
  {Kozuma}, {Sugiura}, {Kuga}, {Deng}, and {Hagley}}}]{torii:2000}
\bibinfo{author}{\bibfnamefont{Y.}~\bibnamefont{{Torii}}},
  \bibinfo{author}{\bibfnamefont{Y.}~\bibnamefont{{Suzuki}}},
  \bibinfo{author}{\bibfnamefont{M.}~\bibnamefont{{Kozuma}}},
  \bibinfo{author}{\bibfnamefont{T.}~\bibnamefont{{Sugiura}}},
  \bibinfo{author}{\bibfnamefont{T.}~\bibnamefont{{Kuga}}},
  \bibinfo{author}{\bibfnamefont{L.}~\bibnamefont{{Deng}}}, \bibnamefont{and}
  \bibinfo{author}{\bibfnamefont{E.~W.} \bibnamefont{{Hagley}}},
  \bibinfo{journal}{\pra} \textbf{\bibinfo{volume}{61}},
  \bibinfo{pages}{041602} (\bibinfo{year}{2000}).

\end{thebibliography}

\end{document}